\documentclass[prl,twocolumn,showpacs,superscriptaddress]{revtex4-1}
\usepackage{graphicx}
\usepackage{url}

\begin{document}

\title{Anisotropic antiferromagnetic order in spin-orbit coupled trigonal lattice Ca$_2$Sr$_2$IrO$_6$}

\author{Jieming Sheng}
\affiliation{Department of Physics, Renmin University of China, Beijing 100872, China}
\affiliation{Neutron Scattering Division, Oak Ridge National Laboratory, Oak Ridge, Tennessee 37831, USA}
\author{Feng Ye}
\email{yef1@ornl.gov}
\author{Christina Hoffmann}
\affiliation{Neutron Scattering Division, Oak Ridge National Laboratory, Oak Ridge, Tennessee 37831, USA}
\author{Valentino R. Cooper}
\author{Satoshi~Okamoto}
\affiliation{Materials Science and Technology Division, Oak Ridge National Laboratory, Oak Ridge, Tennessee 37831, USA}
\author{Jasminka~Terzic}
\altaffiliation{Now at National High Magnetic Field Laboratory, Tallahassee, FL 32306, USA}
\affiliation{Department of Physics, University of Colorado at Boulder, Boulder, Colorado 80309, USA}
\author{Hao~Zheng}
\author{Hengdi~Zhao}
\author{G.~Cao}
\affiliation{Department of Physics, University of Colorado at Boulder, Boulder, Colorado 80309, USA}
\date{\today}

\begin{abstract}
    We used single-crystal x-ray and neutron diffraction to investigate the crystal
    and magnetic structures of trigonal lattice iridate $\rm Ca_2Sr_2IrO_6$.
    The crystal structure is determined to be $R\bar3$ with two distinct Ir
    sites.  The system exhibits long-range antiferromagnetic order below $T_N
    = 13.1$ K. The magnetic wave vector is identified as $(0,0.5,1)$ with
    ferromagnetic coupling along the $a$ axis and antiferromagnetic
    correlation along the $b$ axis. Spins align dominantly within the basal
    plane along the [1,2,0] direction and tilt 34$^\circ$ towards the $c$ axis.
    The ordered moment is 0.66(3) $\mu_B$/Ir, larger than other iridates where
    iridium ions form corner- or edge-sharing $\rm IrO_6$ octahedral networks.
    The tilting angle is reduced to $\approx19^\circ$ when a magnetic field
    of 4.9 Tesla is applied along the $c$ axis. Density functional theory
    calculations confirm that the experimentally determined magnetic
    configuration is the most probable ground state with an insulating gap
    $\sim0.5$~eV.
\end{abstract}

\maketitle

\section{Introduction}
Controlling the balance between spin-orbit interactions (SOI), on-site Coulomb
interactions and crystalline electric field splitting in 5$d$ iridates is
the central theme behind searching for novel quantum phenomena such as $j_{\rm
eff}=$1/2 Mott insulating states \cite{Kim08,Moon08,Kim09}, correlated
topological insulators \cite{Pesin10,Shitade09}, spin liquid phases
\cite{Okamoto07}, superconductivity \cite{Wang11,Watanabe13}, and Kitaev
models \cite{Jackeli09,Price12,Singh12}. Due to the entangled spin and
orbital degrees of freedom, the form of magnetic interactions is no longer
dictated by a global spin SU(2) symmetry. This leads to physics that is
dramatically different from the 3$d$ systems where SOI is of a perturbative nature.
The wave functions are composed by the superposition of different orbital and
spin states and the resulting magnetic interactions depend critically on the
lattice symmetry. In the case of a 180$^\circ$ Ir-O-Ir bond, the
Hamiltonian is governed by an isotropic Heisenberg term plus a weak
dipolar-like anisotropy term due to Hund's coupling, while for a
90$^\circ$ bond, the anisotropic term due to the off-diagonal hopping matrix
results in a quantum analog of the compass model \cite{Jackeli09}. The strong
SOI limit also assumes local cubic symmetry of the IrO$_6$ octahedra, which is
rare in real materials. It was discovered that nearly all iridate families
have a certain degree of noncubic distortions. For example, the O-Ir-O bond
angle in pyrochlores $R\rm{_2Ir_2O_7}$ ($R$ denotes rare earth) is an average
6$^\circ$-10$^\circ$ away from 90$^\circ$\cite{Taira01}, a substantial
elongation (tetragonal distortion) of the $\rm IrO_6$ octahedra occurs in the
$j_{\rm eff}=1/2$ Mott insulator $\rm Sr_2IrO_4$ \cite{Crawford94}, and an
appreciable trigonal distortion was revealed in a honeycomb lattice $\rm
Na_2IrO_3$ with O-Ir-O bond angles $\sim 4-9^\circ$ deviating from the cubic
case \cite{Choi12,Ye12Na2IrO3}. Although it was claimed that the $j_{\rm eff}
= 1/2$ state is robust against distortion, recent resonant inelastic x-ray
scattering (RIXS) studies have shown that the distortion of IrO$_6$ octahedra
leads to a modification of the isotropic wave functions \cite{Liu12,Sala14}.
This underscores the need for extending the $j_{\rm eff}=1/2$ picture to
correctly describe the Mott insulating ground states.

In this paper, we report a single crystal x-ray and neutron diffraction
investigation of a trigonal lattice iridate Ca$_2$Sr$_2$IrO$_6$ (CSIO).
The crystal orders antiferromagnetically (AFM) below 13.1~K with no
structural anomaly across the transition. The wave vector of the spin structure
is (0,0.5,1) indicating strong anisotropic magnetic interactions. The
iridium moments align nearly along the diagonal O-Ir-O direction within the IrO$_6$
octahedra. The ordered moment reaches 0.66(3)$\mu_B$/Ir, larger than other
iridates which form corner- or edge sharing $\rm IrO_6$ octahedral networks.
Most importantly, the local environment of $\rm IrO_6$ is close to the cubic limit
and there is no direct connectivity between individual IrO$_6$ octahedra,
making this system a canonical candidate to study the novel magnetism arising
from the SOI.

\section{Experimental Results}
Single crystals of CSIO were grown using a self-flux method similar to the one
reported in Ref.~[\onlinecite{Cao07b}], from off-stoichiometric quantities of
IrO$_2$, $\rm CaCO_3$ and $\rm SrCO_3$ that were mixed with $\rm CaCl_2$
and/or $\rm SrCl_2$. The starting ratio of Ir to (Ca,Sr) is approximately 1:5.
The mixed powders were fired to 1460$^\circ$C for 4 hours and then slowly
cooled at a rate of 4$^\circ$C/hour. The compositions were
independently checked to be consistent using both energy dispersive x-ray
analysis (EDX) (Hitachi/Oxford 3030 Plus) and single-crystal x-ray
diffraction.  The magnetic susceptibility and specific heat were measured
using a Quantum Design Magnetic Property Measurement System. X-ray diffraction
data were collected using a Rigaku XtaLAB PRO diffractometer at the Oak Ridge
National Laboratory (ORNL). A molybdenum anode was use to generate x ray with
wavelength $\lambda=0.7107$~\AA.  Neutron diffraction measurement was carried
out using the TOPAZ diffractometer with a crystal size of 1$\times$ 1 $\times$
1.5 mm$^3$ at the Spallation Neutron Source (SNS), ORNL. A larger piece with
dimensions 1.5$\times$ 1.5 $\times$ 4 mm$^3$ was chosen for magnetic
structure determination using the single-crystal diffuse scattering
diffractometer CORELLI at SNS \cite{Ye18Corelli}.  The $\pm$ 28.5$^\circ$ vertical
angular coverage of the detector allows an extensive survey in reciprocal
space. A 5~Tesla vertical field superconducting magnet was used to study the
field evolution of the spin structure.

Pure and Sr-doped $\rm Ca_4IrO_6$ were reported to crystallize in a
rhombohedral, K$_4$CdCl$_6$-type structure with $R\bar{3}c$ space group (SG
No.~167) from x-ray powder diffraction studies \cite{Segal96}. The lattice
parameters increase monotonically with Sr doping. The values become $a = b =
9.588$~\AA~and $c = 11.414$~\AA~for CSIO at room temperature. The crystal
structure in Fig.~1(a) shows the one-dimensional (1D) chains of alternating
IrO$_6$ octahedra and CaO$_6$ trigonal prisms parallel to the $c$ axis.  The
single crystal x-ray diffraction measurement on CSIO reveals that the majority
of reflections are consistent with the reported $R\bar3c$ space group, with a
significant portion of peaks violating the reflection conditions [345 out of
3152 reflections with $I>3\sigma(I)$]. To further confirm the finding, we
employed single crystal neutron diffraction to characterize the structure.
Figure~1(b) presents a typical contour plot in the $(h,k,l=1)$ scattering plane.
Indeed, several marked reflections cannot be indexed using SG 167, which
requires both $h+l=3n$ and $l=2n$ in the $(h\bar h0l)$ scattering plane.  The
presence of (4,0,1) in Fig.~1(b) clearly indicates the breakdown of the
reflection condition and suggests a reduced crystal structure symmetry.  Based
on the x-ray and neutron observation, the maximal non-isomorphic subgroup
$R\bar{3}$ (No.~148) that lacks the $c$-axis glide is the most likely space
group, where the unique Ir site ($6b$ site in SG 167) splits into $3a$ and
$3b$ Wyckoff positions. Such different surrounding oxygen environments allow
the two Ir sites to have independent spin orientations. Each $\rm IrO_6$
octahedron contains six identical Ir-O bond length of 2.036 \AA. There is a
small trigonal distortion with the octahedron stretched along the $c$ axis;
the corresponding O-Ir1-O and O-Ir2-O bond angles at 100 K are 88.92(8)$^\circ$
and 88.68(8)$^\circ$, respectively. The overall local environment surrounding
the Ir atoms is close to the ideal cubic limit.  Furthermore, the alkaline
earth atoms connecting neighboring $\rm IrO_6$ along the $c$ axis are
dominated by Ca ions while the sites between the $\rm IrO_6$/$\rm CaO_6$
chains have mixed Sr:Ca ions with a 2:1 ratio [Fig.~1(a)]. This atomic
arrangement is probably due to the longer distance from the mixed site to the
oxygen atoms, which is more suitable to host the larger Sr ions. This feature also
agrees with a 2.7\% increase in $a$, but only 1.6\% increase in $c$ from
$\rm Ca_4IrO_6$ to CSIO. It is noteworthy that the formation of two distinct
octahedral sites is not common in the $\rm K_4CdCl_6$-type compounds. The
observation of the $R\bar{3}$ space group in CSIO might be related to the
preferred site occupancy of the mixed Ca/Sr ions. It is certainly interesting
to verify whether this is the case in similar material with the $R\bar{3}$ space group
discovered in the future.

Figure~1(c) shows the $T$ dependence of the specific heat of CSIO single crystal.
A sharp anomaly appears near 13 K indicating a magnetic transition similar to
pure $\rm Ca_4IrO_6$ \cite{Cao07b,Calder14}. Fig.~1(d) shows the temperature evolution of the
magnetic susceptibility $\chi$  with the applied magnetic field of 0.5~$T$. The
peaks observed around 13.5 K confirm the phase transition. Fits of
$1/\chi_{ab}$ ($1/\chi_{c}$) for  30~K $<T<$300~K to a Curie-Weiss law yield
effective magnetic moments $\mu_{\rm eff}$ of 1.25 (2.14)$\mu_B$ and Curie
temperatures $\theta_{\rm CW}$ of -0.13 (-12.69) K for the field applied in the
basal plane (parallel to the $c$ axis). The negative value in $\theta_{\rm CW}$
implies AFM interactions between the neighboring Ir ions. The large difference
in both $\mu_{\rm eff}$ and $\theta_{\rm CW}$ within the $ab$ plane and along
the $c$ axis indicates strong anisotropy in magnetic property and is
consistent with the chain-like topology of the crystal structure. The average
value ($2/3\theta ^{ab}_{\rm CW}+1/3\theta^c_{\rm CW}$) agrees well with the powder
sample, where $\theta_{\rm CW}$ decreases steadily with Sr doping \cite{Segal96}. For
systems with a trigonal or triangular lattice, it is generally expected there is
a certain amount of magnetic frustration \cite{Ramirez94}.  However, the small
value of frustration parameter ($\theta_{\rm CW}/{T_N}$ $\approx$ 0.32) implies it
is absent in CSIO.

\begin{figure}[ht!]
 \includegraphics[width=3.4in]{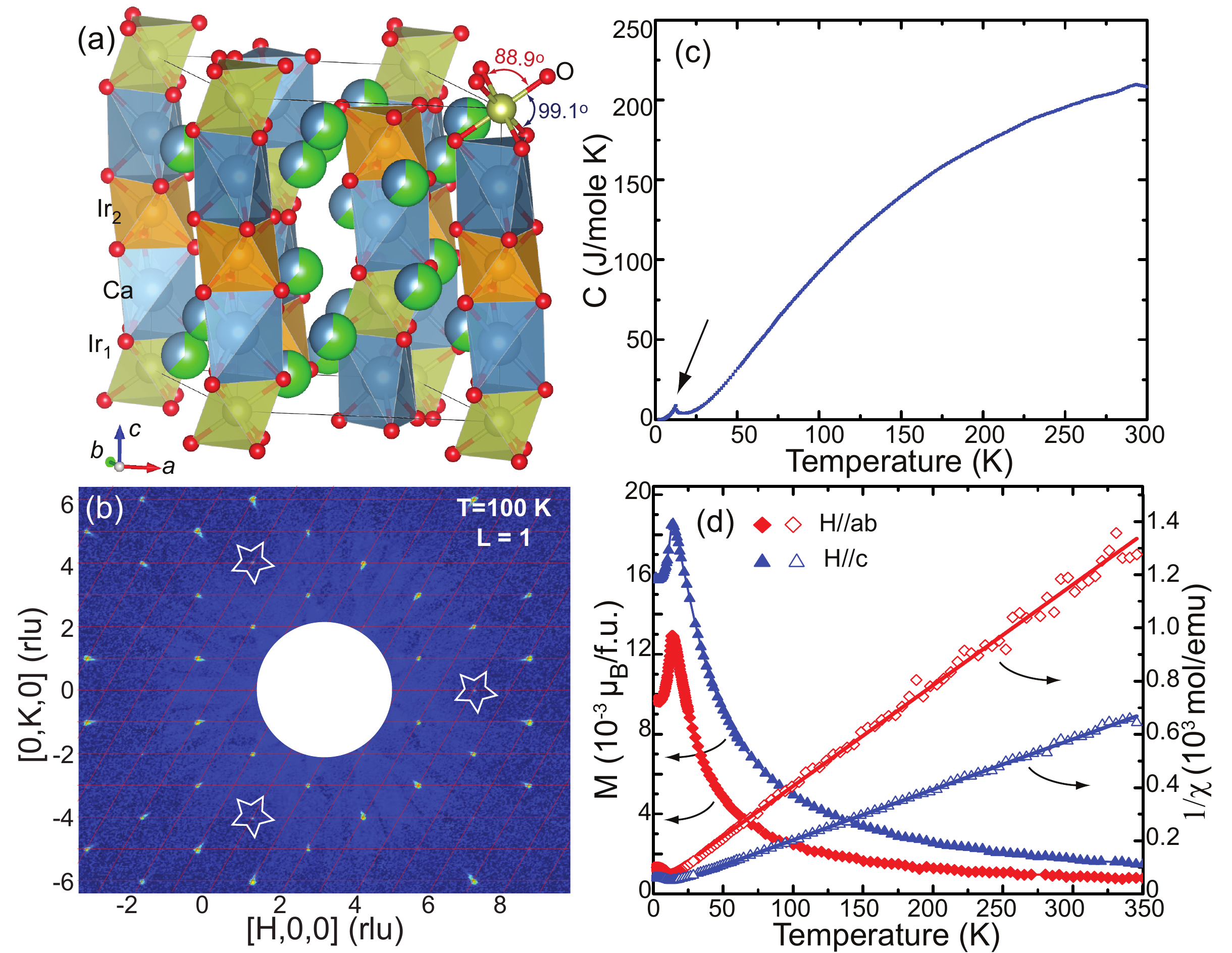}
    \caption{(a) Crystal structure of $\rm Ca_2Sr_2IrO_6$ in SG $R\bar3$.
    There are two distinct Ir1 and Ir2 sites located at
    (0,0,0) and (0,0,0.5) with different bonding oxygen environment. The
    trigonal distortion of $\rm IrO_6$ leads to out-of-plane 88.9$^\circ$
    and in-plane 91.1$^\circ$ bond angles.  The structure is drawn using
    {\footnotesize VESTA} software \cite{Momma11}. (b) The reciprocal-space
    image in the $(h,k,l=1)$ scattering plane at 100~K with data collected
    from the TOPAZ diffractometer. The nuclear peaks in white stars are
    forbidden reflections of SG $R\bar3c$. (c) Temperature dependence of
    specific heat $C_p(T)$. (d) $T$-dependence of magnetization $M(T)$ and
    inverse magnetic susceptibility $1/\chi(T)$ in an applied magnetic field of
    0.5 T parallel to the $ab$ plane and the $c$ axis in the field cooling
    protocol. Solid lines are the fits using Curie-Weiss law above transition
    temperature.}
\end{figure}

The spin structure of CSIO in a zero applied magnetic field ($\rm H=0$) was
characterized by surveying a large portion of the reciprocal volume at 5 K.
The sample was oriented with the $c$ axis perpendicular to the horizontal
scattering plane.  The diffraction data were collected with the crystal
rotating along the $c$ axis for 210 degrees. Earlier studies of undoped
$\rm Ca_4IrO_6$ reported a spin configuration with a magnetic wave vector
$q_m=(0.5,0.5,0)$ \cite{Calder14}.  As shown in Fig.~2(a), the low-$T$ contour
plot in the $(h,k,l=0)$ scattering plane does not show extra intensities at
this reflection and equivalent positions. In contrast, new reflections appear
in the plane with $l=2n+1$. All observed magnetic reflections at $(h,k,l=1)$
can be indexed using a magnetic wave vector $q_m=(0, 0.5,1)$ plus two
additional magnetic domains  -120 and 120 degrees apart [Fig.~2(b)]. The
volume fraction ratio of the three magnetic domains is 35:33:32, and is
consistent with the trigonal symmetry. The observed magnetic propagation
wave vector $(0, 0.5, 1)$ in CSOI is the same as the isostructural $\rm
Sr_3ZnIrO_6$ \cite{McClarty16}. The $T$ dependence of the strongest
magnetic reflection shows a clear second order phase transition.  Fitting the
data using $I \sim (1-T/T_N)^{2\beta}$ yields $T_N = 13.1(3)$~K and $\beta =
0.25(1)$. The value of $\beta$ deviates from the critical exponent of a 3$D$
spin system but is consistent with pure $\rm Ca_4IrO_6$ \cite{Calder14}.

\begin{figure}[thb!]
     \includegraphics[width=3.4in]{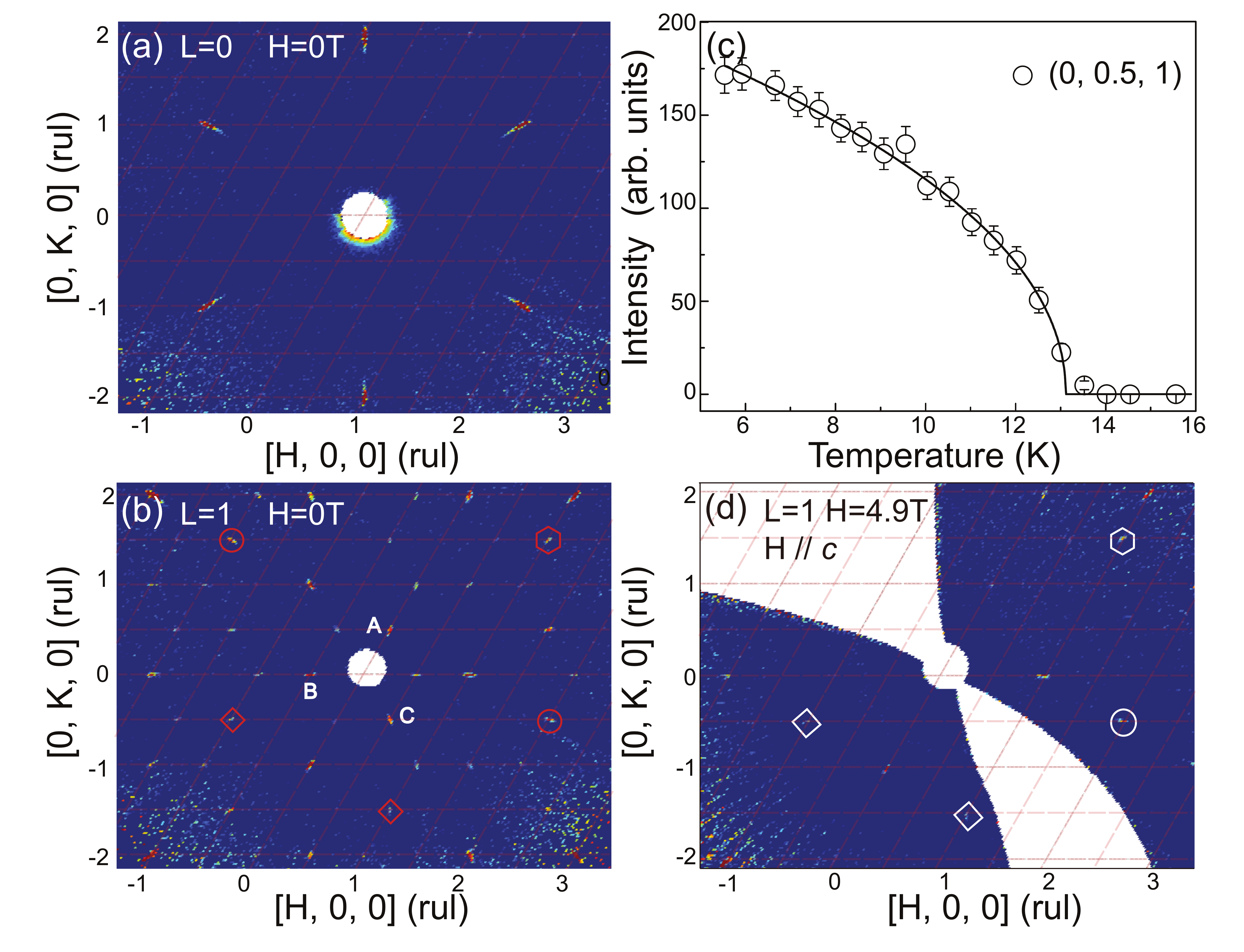}
     \caption{The diffraction image in the (a) $(h,k,l=0)$ and (b) $(h,k,l=1)$
     scattering plane at 5 K with zero field. Capital letters $A$, $B$, $C$ label
     the three magnetic domains.  Reflections encircled in squares, circles and
     hexagons are from the same magnetic domain with the wave vector
     $q_m=(0,0.5,1)$. (c) $T$ dependence of the (0,0.5,1) magnetic peak. The solid
     line is the fit of integrated intensity. (d) The image in the $(h,k,l=1)$
     scattering plane with a magnetic field of $H=4.9$~T applied along the
     $c$ axis. The marked peaks are the magnetic reflections.}
\end{figure}

The magnetic structure of CSIO is determined by
analyzing over 80 reflections in conjunction with representational
analysis \cite{Rodriguezcarvajal93}. For the SG $R\bar3$ with two inequivalent
Ir sites and propagation wave vector $(0,0.5,1)$, the spin
configuration is described by an irreducible representation (IR)  that allows
moments of Ir1(Ir2) along all three crystallographic axes. The relative phase
between the two iridium sites could either be ferromagnetic (FM) or AFM. In
the former case, one expects strong magnetic reflections in the scattering
plane with $l$ equal to even numbers.  However, all major magnetic peaks are
observed in the scattering plane with $l$ equal to odd numbers, indicating a
dominant AFM coupling between the two iridium sites. This feature is verified
using the simulated annealing method \cite{Rodriguezcarvajal93}.
There are weak peaks in the $l=0$ plane, e.g., the $(2,0.5,0)$ peak. This
indicates that the spins at those two sites are not exactly out of phase
and further confirms the reduced crystal symmetry. A reliable refinement
of the spin configuration can only be reached by collecting a complete set of
magnetic reflections in reciprocal space with domain populations correctly
refined, since certain reflections result from the summation of different
domains. Full details are given in the Supplemental Material \cite{CSIOSM}.
Using the symmetry-adapted model and the magnetic form factor for Ir$^{4+}$
\cite{Kobayashi11}, we obtained the spin structure with detailed information
listed in Table~I.  As illustrated in Figs.~3(a)-3(b), the spins have
staggered $+-$ patterns between neighboring sites along the $c$ axis. The
in-plane spin configuration is highly anisotropic despite the trigonal
symmetry of the lattice. The moments are coupled ferromagnetically along the
$a$ axis but are antiferromagnetic along the $b$ axis. The spin moments are
dominantly aligned within the $ab$ plane with an out-of-plane tilt angle of
34$^\circ$ towards the $c$ axis.  The projection in the basal plane is
parallel to the [1,2,0] direction. The moment direction is close to the
diagonal O-Ir-O bond within the IrO$_6$. The nearly collinear spin
configuration is in contrast with the noncollinear spin order reported in the
isostructural $\rm Sr_3ZnIrO_6$ \cite{McClarty16}. The ordered moment in CSIO
is 0.66(3)$\mu_B$/Ir site. It is larger than other iridates compounds with
corner- or edge sharing IrO$_6$ octahedra
\cite{Cao00,Lovesey12b,Ye12Na2IrO3,Ye13Sr2IrO4,Dhital12a,Dhital13}, but
comparable with the value of 0.87 $\mu_B$/Ir in $\rm Sr_3ZnIrO_6$ and 0.6
$\mu_B$/Ir in the transition metal element substituted $\rm Sr_3CoIrO_6$ that has
a similar crystal structure with quasi-1$D$ chains along the $c$ axis
\cite{McClarty16,Mikhailova12}.  Thus, the relative large ordered moment is
most likely due to the suppressed electron hopping between the isolated
IrO$_6$ octahedra.

\begin{table}[htb!]
    \caption{The basis vectors (BV) and refined spin components for the SG $R\bar3$
    with $q_m = (0,0.5,1)$. Two independent Ir sites are located at (0,0,0)
    and (0,0,1/2), respectively.}
\begin{ruledtabular}
\begin{tabular}{lccccc}
    IR     & BV & atom    & $m_a$ & $m_b$ & $m_c$\\
       $\Gamma_1$  & $\psi_1$  &  Ir1 &   1   &  0    &   0\\
                   & $\psi_2$  &  Ir1 &   0   &  1    &   0\\
                   & $\psi_3$  &  Ir1 &   0   &  0    &   1\\
                   \hline
       $\Gamma_1$  & $\psi_1$  &  Ir2 &   1   &  0    &   0\\
                   & $\psi_2$  &  Ir2 &   0   &  1    &   0\\
                   & $\psi_3$  &  Ir2 &   0   &  0    &   1\\
                   \hline
    \multicolumn{3}{r}{Refinement ($T = 5$K)}  & $m_a ({\rm \mu_B})$ & $m_b({\rm \mu_B})$ & $m_c({\rm \mu_B})$\\ \hline
                   &           &  Ir1 &   -0.33(2)   &  -0.64(2)    & 0.41(5)\\
                   &           &  Ir2 &    0.32(2)   &   0.63(2)    &-0.34(5)\\
\end{tabular}
\end{ruledtabular}
\end{table}

The magnetic-field effect on the spin structure is investigated with
a field applied along the $c$ axis. The diffraction pattern in the $(h,k,l=1)$
scattering plane is presented in Fig.~2(d). Compared to the zero field data,
only one of the three magnetic domains survives at $H=4.9$~T. This supports
the conclusion that the observed magnetic reflections at $H=0$ result from
multiple domains instead of one single magnetic domain with multi-$k$
structure \cite{Gallego16}.  Limited number of magnetic reflections are
collected due to the partial block of a neutron beam by the magnet. With the
spin configuration constrained to be similar to that at $H=0$, the refined spins
tilt further towards to the basal plane with a canting angle of 19$^\circ$, and the
ordered magnetic moment decreases to 0.60(7)$\mu_B$/Ir [Fig.~3(d)]. The
tilting of the moment direction towards to the basal plane is expected since
the state with field $H$ perpendicular to the easy magnetization is
energetically more favorable. On the other hand, the absence of a spin-flip
transition with field up to 4.9 ~T (the transition temperature $T_N$ reduces
from 12.5~K at $H=0$ to 10.6~K at $H=14$~T from specific heat measurement)
indicates the magnetic structure is rather robust and consistent with
observations in other iridates \cite{Ye12Na2IrO3}.

\begin{figure}[thb!]
     \includegraphics[width=3.4in]{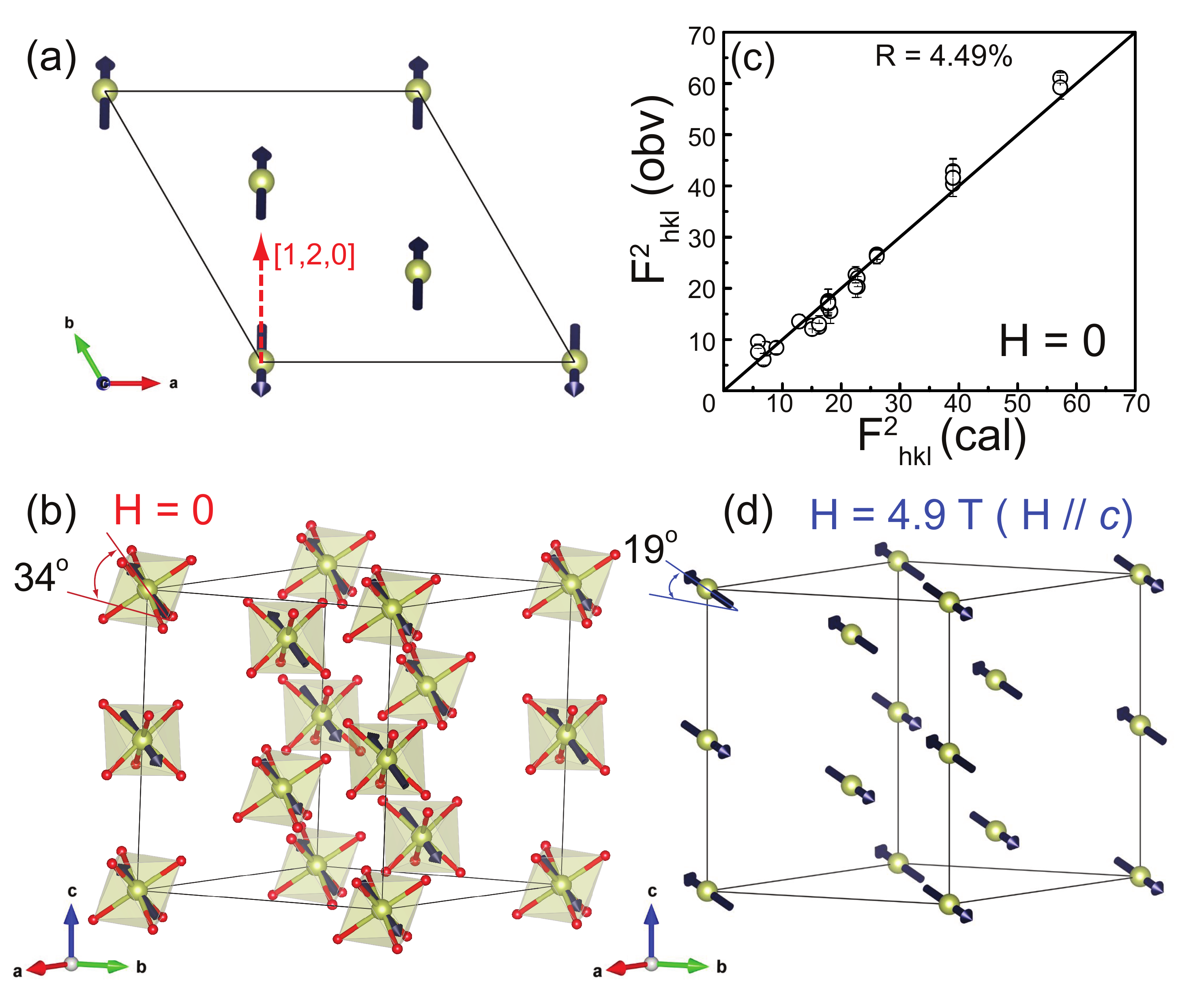}
     \caption{(a) The spin configuration projected on the $ab$ plane with
     moment along the [1,2,0] direction. (b) The refined magnetic structure at
     $H=0$ from single-crystal neutron diffraction measurement. the spin
     moments tilt 34$^\circ$ away from the $ab$ plane. (c) The calculated magnetic
     intensities versus observation using the model described in the text. (d)
     The magnetic structure with $H=4.9$~T applied along the $c$ axis; the
     canting angle away from the basal plane reduces to 19$^\circ$.  }
\end{figure}

To further understand the nature of the observed magnetic ordering, we performed
density functional theory (DFT) calculations using the Vienna Ab initio
Simulation Package {\small VASP} \cite{Kresse96a,Kresse96b} with the modified
Perdew-Burke-Erznenhoff exchange-correlation designed for solids (PBEsol)
\cite{Perdew08}. We employed PAW potentials \cite{Blochl94} with the following
electronic configurations: Ca:$3p^63s^2$, Sr:$4s^24p^65s^2$, Ir:$6s^15d^8$, and
O:$2s^22p^4$. The cation arrangement was chosen using a random number
generator to assign Ca or Sr with the correct distribution on each site.  The
calculations were found to be converged with a 500~eV
cutoff. To allow for the two antiferromagnetic configurations to be studied in
a commensurate unit cell, $2\times2\times1$ unit cells were employed with a
$1\times1\times2$ Monkhorst-Pack $k$-point mesh. All ionic coordinates were
relaxed until all Hellman-Feynman forces were less than 0.015~$\rm eV/\AA$. A
Hubbard $U$ of 2.0~eV and intrasite Hund's coupling $J_H=0.2$~eV for Ir
$d$-states were employed \cite{Dudarev98}. The magnetic structure shown in
Fig.~3 (AFM3 state) and the one reported for pure $\rm Ca_4IrO_6$ (AFM1 state
with $q_m=(0.5,0.5,0)$ as shown in Ref.~[\onlinecite{Calder14}]) were chosen
as the initial magnetic configurations. With SOI taken into account, the
density of states (DOS) for both magnetic structures exhibit similar
features. However, the energy of the AFM3 state is 2.5 meV/Ir site lower than
that of AFM1 state, thereby confirms the observed magnetic structure as the
most probable ground state. The initial magnetic structure tested in the
calculation has the moments direction arbitrarily chosen while keeping the
configuration consistent with the magnetic wave vector, but the converged state
has moments relaxed nearly along the Ir-O bond direction.  There is an
insulating gap $\sim 0.5$~eV near $E_F$, mainly from the $t_{2g}$ orbital of
the Ir$^{5+}$ ions.  This is consistent with the resistivity measurement shown in
Fig.~4(b) where an band gap (2$\Delta_{ag}$) $\sim$ 1.26~eV is obtained by
fitting the data to the form of $\rho(T) = \rho_0 \exp(\Delta_{ag}/k_BT)$ for
300 $<T<$ 500~K. The calculated spin moment of 0.5$\mu_B$/Ir agrees well
with the experimental observation.

\begin{figure}[thb!]
     \includegraphics[width=3.4in]{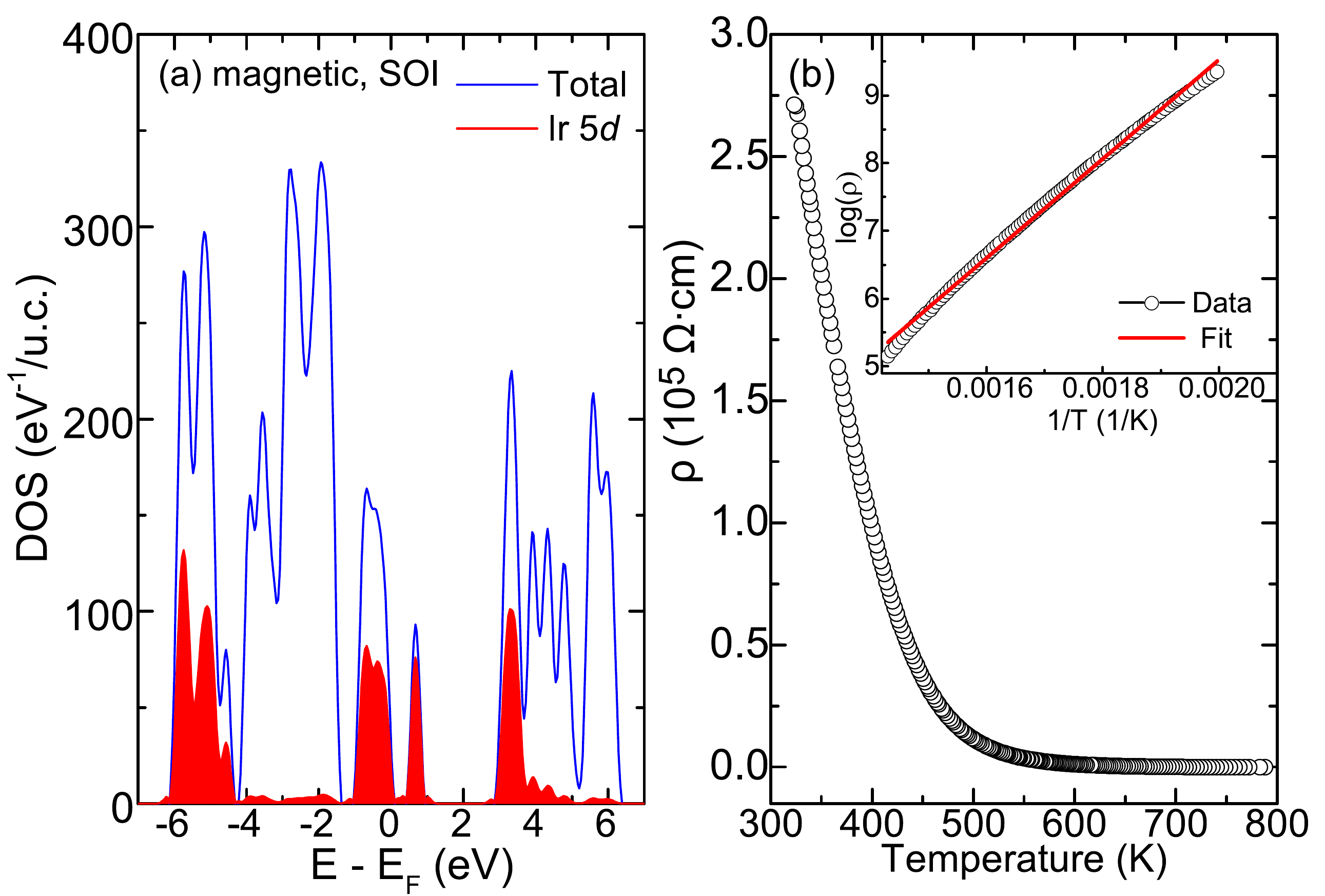}
     \caption{ The density of states of CSIO from DFT calculation with
     spin-orbit interaction included. The initial magnetic structure is
     similar to the one illustrated in Fig.~3.  (b) The temperature
     dependence of resistivity of CSIO.  Inset shows the fit to $\rho(T) =
     \rho_0 \exp(\Delta_{ag}/k_BT)$ with $\Delta_{ag}=0.63$~eV.}
\end{figure}

\section{Discussion}

The anisotropic magnetic configuration that breaks the trigonal symmetry and
the moment direction following the diagonal O-Ir-O bond strongly suggests the
emergence of relativistic SOI, where the exact form of the magnetic
Hamiltonian depends on the lattice geometry \cite{Jackeli09}. A Heisenberg
interaction $\vec{S}_i\cdot\vec{S}_j$ dominates in a corner-shared iridate,
e.g.,~the square lattice $\rm Sr_2IrO_4$. The canted spin moments that
rigidly follow the staggered rotation of octahedra
\cite{Ye13Sr2IrO4,Boseggia13b} are naturally explained by the strong SOI.
In contrast, the highly anisotropic interactions appear due to the
off-diagonal hopping matrix in the edge-shared case. This maps the system into
a quantum compass model and has been studied extensively in the honeycomb
lattice $\rm A_2IrO_3$ \cite{Chaloupka13,Rau14a}. Since the building blocks of
perovskite-derived iridates are made of individual IrO$_6$ octahedrons, the
isolated $\rm IrO_6$ without corner- or edge-sharing connectivity,
makes the CSIO an ideal system to study the SOI in the single-ion limit. On the
other hand, the influence of nonoctahedral crystal-field splitting ($\Delta$)
can not be ignored.  Although the $ j_{\rm eff}=1/2$ state was initially
thought to be a robust feature in iridates, as evidenced by the vanishing
intensity at the $L_2$ absorption edge \cite{Kim09}, it is now recognized that
the local distortion could dramatically modify the ground states. For example,
RIXS measurements on a quasi-one-dimensional spin-chain $\rm Sr_3CuIrO_6$ have
revealed $\Delta=0.31$~eV caused by the reduction of the O-Ir-O bond angle to
82$^{\circ}$, and they contributed a significant mixing between $ j_{\rm
eff}=1/2$ and $j_{\rm eff}=3/2$ states \cite{Liu12}.  A similar
result is reported in post perovskite $\rm CaIrO_3$ \cite{Sala14}, where
the energy scale of the octahedral compression along the local $z$ axis
($\Delta=-0.71$~meV) is comparable with the SOI strength $\lambda=0.52$~meV and
signifies a departure from the $ j_{\rm eff}=1/2$ state.
While the local symmetry of these compounds is not the same, i.e., tetragonal for
perovskites and trigonal for non-perovskites, distortions of
the IrO$_6$ octahedron seem to be ubiquitously present in the so-called $j_{\rm
eff}=1/2$ iridates (e.g.~$\rm Na_2IrO_3$ with $|\Delta|=0.11$ eV
\cite{Gretarsson13a}, $\rm Y_2Ir_2O_7$ with $|\Delta|$=0.59 eV
\cite{Hozoi14}). A nonoctahedral crystal field must be considered in realistic
models since the electronic structure is highly dependent on the relative orbital
contributions. In this respect, the $A_4B{\rm O_6}$ ($A$ denotes alkaline earth
ions and $B$ is a 4$d$ or 5$d$ element) system featuring a chain-like structure
with minimal local distortion of $B$O$_6$ octahedra represents a new family of
platforms to realize the spin-orbit-entangled state. Yet, distinct magnetic
configurations have been reported in isostructural iridates such as spin order
with wave vector (0.5, 0.5, 0) in $\rm Ca_4IrO_6$ \cite{Calder14}, or a
noncollinear spin structure in $\rm Sr_3ZnIrO_6$ \cite{McClarty16}. The
difference indicates that the magnetic coupling between seemingly isolated Ir
octahedra might depend on the overall averaged lattice and warrant more
experimental investigation. Unlike the transition metal
substituted $\rm Sr_3NiIrO_6$ or $\rm Sr_3CoIrO_6$ where the magnetism is
influenced by the interplay between the transition metal and the $5d$ ions, or the
$4d$ counterpart $\rm Sr_4RhO_6$ \cite{Calder15b} where the strength of SOI is
smaller than the 5$d$ systems, the CSIO can be regarded as a suitable example
to further explore the electronic and magnetic properties arising from the
SOI.

In summary, neutron and x-ray diffraction have been employed to investigate
the crystal and magnetic structures of the trigonal lattice iridate $\rm
Ca_2Sr_2IrO_6$. The well-separated IrO$_6$ octahedra are close to the cubic limit
with six equal Ir-O bond distance and O-Ir-O bond angles $\approx90^\circ$.
The Ir$^{4+}$ spins form an anisotropic three-dimensional antiferromagnetic
configuration with wave vector (0,0.5,1). The ordered moment is 0.66(3)$\mu_B$/Ir, 
which is larger than iridates with corner- and edge-sharing $\rm IrO_6$
octahedral networks. The DFT calculation confirms that the observed magnetic
ordering is the most probable ground state and indicates that the insulating
behavior is enhanced by the spin-orbit interaction.

We thank G. Jackeli and J.M. Perez-Mato for stimulating discussion. Research
at ORNL's SNS was sponsored by the Scientific User
Facilities Division,  Basic Energy Sciences, U.S.~Department of
Energy (DOE). Theoretical calculations were supported by the Materials
Sciences and Engineering Division (SO), Office of Basic Energy Sciences, U.S.
DOE, and through the Office of Science Early Career Research Program (VRC).
Computing resources are from NERSC, supported by the office of science,
U.S.~DOE under Contract No.~DE-AC02-05CH11231. Work at Univ.~of Colorado was
supported by the U.S.~National Science Foundation via grant DMR-1712101. JMS
acknowledges support from China Scholarship Council.

This manuscript has been authored by UT-Battelle, LLC
under Contract No. DE-AC05-00OR22725 with the U.S. Department of Energy.  The
United States Government retains and the publisher, by accepting the article
for publication, acknowledges that the United States Government retains a
non-exclusive, paid-up, irrevocable, world-wide license to publish or
reproduce the published form of this manuscript, or allow others to do so, for
United States Government purposes.  The Department of Energy will provide
public access to these results of federally sponsored research in accordance
with the DOE Public Access Plan cite{DOE}.

%
\end{document}